\documentclass[conference]{IEEEtran}
\IEEEoverridecommandlockouts
\usepackage{cite}
\usepackage{amsmath,amssymb,amsfonts}
\usepackage{algorithmic}
\usepackage{graphicx}
\usepackage{textcomp}
\usepackage{xcolor}
\usepackage{stfloats}
\def\BibTeX{{\rm B\kern-.05em{\sc i\kern-.025em b}\kern-.08em
    T\kern-.1667em\lower.7ex\hbox{E}\kern-.125emX}}
\begin{document}

\title{Development Of A Low Cost System For Lifting Related Injury Analysis}

\author{\IEEEauthorblockN{Mahmudul Hasan}
\IEEEauthorblockA{\textit{Technical Service} \\
\textit{Runner Motors Ltd}\\
Tongi, Bangladesh \\
mahmudulhasan290@gmail.com}
\and
\IEEEauthorblockN{Miftahur Rahman}
\IEEEauthorblockA{\textit{School of Aerospace, Transport and Manufacturing} \\
\textit{Cranfield University}\\
Cranfield, UK \\
miftahur.rahman@cranfield.ac.uk}
}

\maketitle

\begin{abstract}
In the developing countries, most of the Manual Material Handling (MMH) related tasks are labor-intensive. It is not possible for these countries to assess injuries during lifting heavy weight, as multi-camera motion capture, force plate and electromyography (EMG) systems are very expensive. In this study, we proposed an easy to use, portable and low cost system, which will help the developing countries to evaluate injuries for their workers. The system consists of two hardware and three software. The joint angle profiles are collected using smartphone camera and Kinovea software. The vertical Ground Reaction Force (GRF) is collected using Wii balance board and Brainblox software. Finally, the musculoskeletal analysis is performed using OpenSim Static Optimization tool to find the muscle force. The system will give us access to a comprehensive biomechanical analysis, from collecting joint angle profiles to generating muscle force profiles. This proposed framework has the potential to assess and prevent MMH related injuries in developing countries. 
\end{abstract}

\begin{IEEEkeywords}
Symmetric lifting, Kinovea, OpenSim, Wii balance board, Static optimization, Muscle force, Ground reaction force, Manual material handling
\end{IEEEkeywords}

\section{Introduction}
Low back pain is one of the leading causes of disability and musculoskeletal disorder (MSDs). The epidemiology of low back pain related disorders got attention in developed countries. The economic impact of low back pain only in the united states is more than \$100 billion \cite{b1}. However, it remains unnoticed in developing countries due to socioeconomic conditions, lack of awareness and low quality healthcare system.\\
Several researchers approached different ways to analyze occupational MSDs. The use of multi-camera motion capture system, force plate and EMG sensors is the traditional approach to evaluate a potential injury. However, these equipment are very expensive, requires technically skilled operators, challenging to set up and transport, making these instruments difficult to use to evaluate occupational injuries. Last few decades, biomechanical human models have become very popular. The most simplified biomechanical model to evaluate lifting related injuries is two-dimensional (2D) skeleton human models \cite{b2}, \cite{b3}, \cite{b4}, \cite{b5}, \cite{b6}. But, for asymmetric lifting, these models are not effective as it misses the differences of kinetics on both sides of the human body \cite{b6}. Three-dimensional (3D) skeleton models can capture the kinetic differences of both sides of the human body during lifting \cite{b7}, \cite{b8}, \cite{b9}, \cite{b10}. Moreover skeletal models cannot give us information about the muscle forces and activations. The injury evaluation using skeletal models may not be accurate, as muscle-tendon dynamics is not counted there. AnyBody (AnyBody technology, Denmark) is a biomechanical modeling software which helps to calculate muscle force, joint reaction forces and moments. But AnyBody Modeling System is not an open-source software.  Most of these biomechanical models and software are lab-based and are not free for all, making it infeasible and inaccessible for the developing countries to use those in industries. OpenSim is a widely popular open-source software for biomechanical modeling, simulation and analysis \cite{b11}. OpenSim is used in evaluating the lifting injuries \cite{b12}, \cite{b13}, \cite{b14}, \cite{b15}, \cite{b16}. But, the input of both Anybody and OpenSim, require kinematics and kinetics data. These require multi-camera human motion capture device and force plate which are very expensive. Several studies were done on Kinovea \cite{b17}, \cite{b18}, \cite{b19}, \cite{b20}, \cite{b21}, \cite{b22} and Wii balance board \cite{b23}, \cite{b24}, \cite{b25}, \cite{b26} as an alternative solution to precision motion capture and force plate, mainly for  measuring the center of pressure and gait analysis. In a study, Wii balance board was integrated with OpenSim to analyze static balance \cite{b27}. No study was done using Kinovea and Wii balance board to analyze lifting related injuries.\\
In this study, we will propose a system by integrating Wii balance board and Kinovea software with OpenSim for musculoskeletal analysis. This low cost system will help the industries of developing countries to perform a comprehensive biomechanical analysis for the workers and assess lifting related injuries. In our knowledge, this is the first study where Kinovea and Will balance board are integrated with OpenSim to evaluate MMH related injuries.  
\section{METHOD}
The low cost biomechanical analysis system consists of three steps: human motion analysis, GRF analysis and musculoskeletal analysis. To evaluate a potential MMH related injury, human kinematics, external forces like ground reaction forces and a muscle force are necessary. We will get anthropometric data of human body and joint angle profile for a lifting motion from human motion analysis part. From GRF analysis step, we will get the vertical GRF. Finally, from the musculoskeletal analysis, we will get the muscle force information.

\subsection{Human Motion Analysis}
For human motion analysis, we used a high-quality camera phone and Kinovea software. The traditional motion capture system is the multi-camera system, which is expensive and difficult to setup. Kinovea is an open-source two-dimensional (2D) motion analysis software.  There is a growing demand for 2D motion analysis software in the biomechanics field, as it reduces complexity and computational cost. Kinovea allows measuring the kinematics of the human body frame by frame. It is easy to use and does not need any physical sensors. The accuracy of the Kinovea software is widely studied \cite{b18}, \cite{b22} and used for different clinical uses \cite{b17}, \cite{b18}, \cite{b19}, \cite{b20}, \cite{b25}.  Deviations in angle estimation between Kinovea and multi-camera system are less than 5 degrees \cite{b25}. The motion video taken from a smartphone will be analyzed in Kinovea. This low cost motion capture system will give us access to generate joint angle profiles during MMH related tasks. As this motion capture system is portable, it is easy to implement in any industrial establishment.

\subsection{Ground reaction force analysis}
Ground reaction force (GRF) is an essential parameter for the biomechanical analysis of the human body. The gold standard for measuring the GRF is the traditional force plate or platform. However, these force plates are very costly, challenging to transport and install. The Wii Balance Board (Nintendo, Kyoto, Japan) is a part of popular video game WiiFit. The accuracy of the Wii balance board has been proven in several studies \cite{b23}, \cite{b25}. The sampling rate of the balance board ranges from 30-50 Hz \cite{b24}, \cite{b25}. We will use the Wii balance board to measure the GRF.

\subsection{Musculoskeletal analysis}
The biomechanical analysis of the lifting motion is performed in OpenSim software. OpenSim is a widely popular   open-source software for biomechanical modeling, simulation and analysis. We will use the static optimization tool of the OpenSim software to find muscle activations and forces.\\
Static optimization is an extension of inverse dynamics. Joint torques can be found from joint kinematics and ground reaction forces using inverse dynamics (Equation 1).
\begin{equation}\label{key}
	M(q)\ddot{q} + C(q, \dot{q}) + G(q) = \tau
\end{equation}
Where, \(\ddot{q} \epsilon R^{3}\) is the angular acceleration, \(M(q) \epsilon R^{3x3}\) is the symmetric positive definite mass (inertia) matrix, \(C(q) \epsilon R^{3x3}\) is the centripetal and coriolis force matrix and \(G(q) \epsilon R^{3}\) is the gravitational force vector.  In static optimization, the joint torque is distributed among the muscles to find the muscle activations and forces. The muscle forces are optimized by minimizing muscle activations (Equation 2).
\begin{equation}
	\begin{gathered}
		minimize: J(a_{m}) \\
		subject \: to: \sum\limits_{m=1}^{n} [a_{m}f(F_{m}^{0},l_{m},v_{m})]r_{m,j}=\tau_{j} 
	\end{gathered}
\end{equation} 
where n is the number of muscles in the model, \(a_{m}\) is the activation level of muscle m at a discrete time step, \(F_{m}^{0}\) is the maximum isometric force, \(l_{m}\) is muscle length, \(v_{m}\) is velocity, \(f(F_{m}^{0},v_{m},l_{m})\) is the muscle force-length-velocity surface, \(r_{m,j}\) is its moment arm about the \(j\textsuperscript{th}\) joint axis, \(\tau_{j}\) is the generalized force acting about the \(j\textsuperscript{th}\) joint axis.
 
\section{DEMONSTRATION OF THE SYSTEM}
The proposed low cost system requires only two hardware and three open-source software. The hardware are smartphone camera and Wii Balance Board. The open-source software are Kinovea, Brainblox and OpenSim. Integration between those devices is shown in Figure 1 as a workflow diagram.
\begin{figure}[ht]
	\centering
	\includegraphics[scale=0.9]{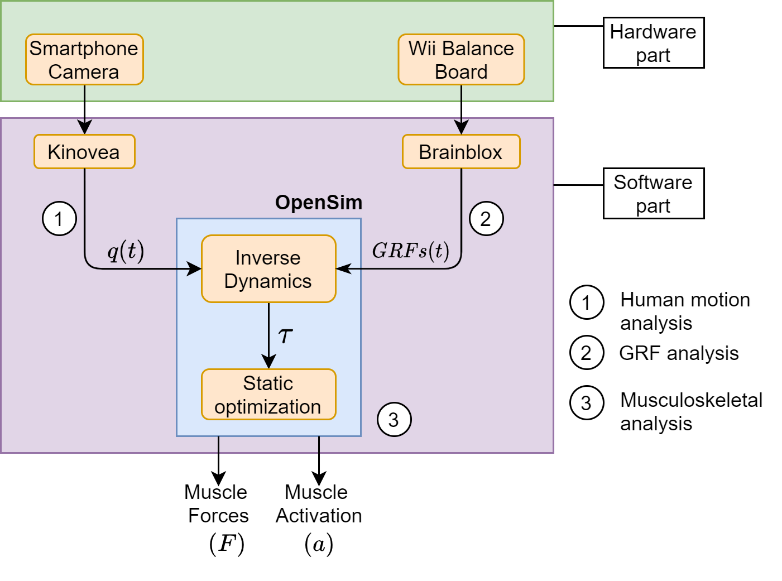}
	\caption{Development of the biomechanical analysis system}
	\label{fig:fig1}
\end{figure}
\\For demonstration purpose, we used a 5-lb (2.26 kg) dumbbell. A healthy volunteer (29 years old, 75 kg) was chosen for the demonstration. The participant signed a written a consent form. The subject did not have any previous physical disorder and had an average physique of south Asian sub-continent. We requested the subject to stand on the Wii balance board and lift the weight from the floor to 1-m height.  The subject repeated the lifting three times. The lifting motions were captured by a smartphone camera. The lifting motion video was analyzed in Kinovea (version 0.9.3) to calculate the joint-angles. The frame rate of lifting motion at Kinovea is 25Hz. We also got the segmental body lengths of the subject at T-position from the Kinovea. Figure 2(a) shows the subject with markers for the weight lifting operation. During the lifting motion, the GRF was collected from the Wii Balance Board (Figure 3). The communication was established between the Wii balance board and a Windows 10 based laptop through Bluetooth. BrainBlox \cite{b28} software was used to collect data from the Bluetooth port.\\
The extracted kinematics and kinetics data were inserted in the OpenSim 4.1 software. A full-body musculoskeletal model \cite{b29} was used for biomechanical analysis. The model has 80 muscle-tendon unit and 17 ideal torque actuators.  The collected kinematics and kinetic data from Kinovea and Wii balance board were inserted in the OpenSim software to calculate muscle force. Figure 2 (b \& c) shows the musculoskeletal model in OpenSim.  
After that, the kinematics and kinetics data were inserted in the OpenSim software to calculate muscle force. 
\begin{figure}[hb]
	\centering
	\includegraphics[scale=0.65]{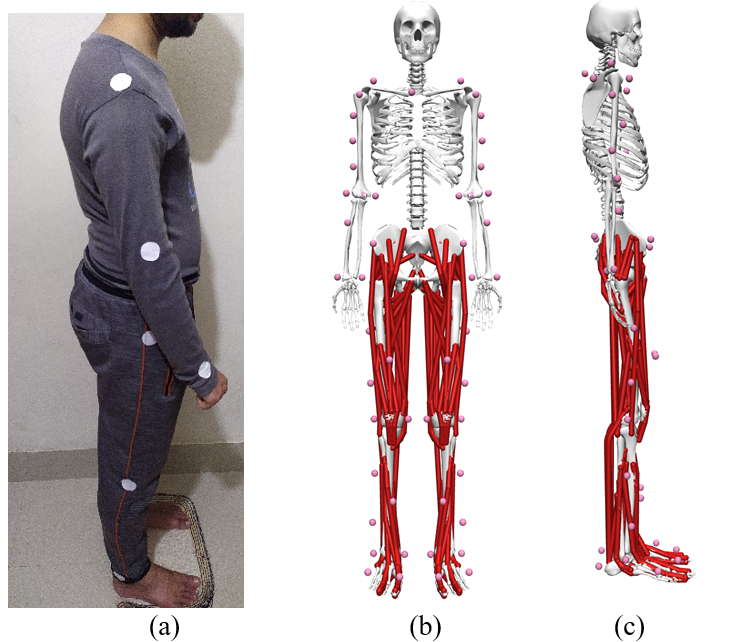}
	\caption{(a) Subject with markers (sagittal plane) (b) musculoskeletal model frontal plane (c) musculoskeletal model (sagittal plane)}
	\label{fig:fig2}
\end{figure}
\begin{figure}[hb]
	\centering
	\includegraphics[scale=0.7]{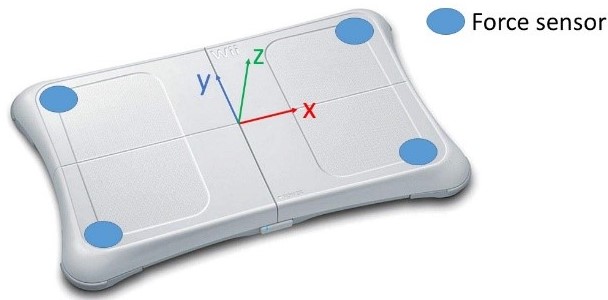}
	\caption{Wii balance board}
	\label{fig:fig3}
\end{figure}
\begin{figure*}[ht]
	\centering
	\includegraphics[scale=0.56]{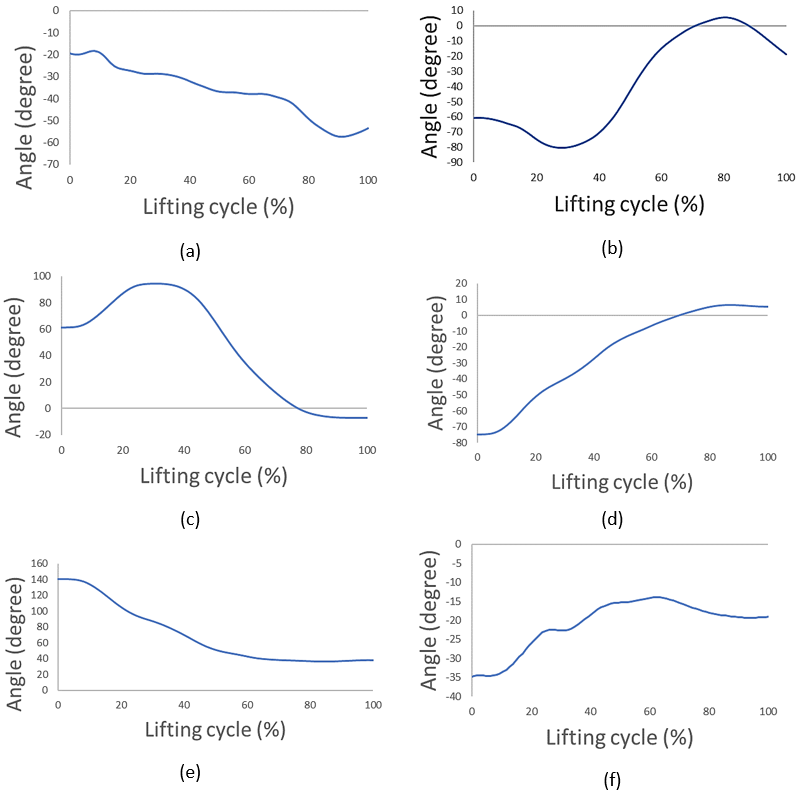}
	\caption{Joint angle profiles of (a) elbow (b) shoulder (c) lumbar spine (d) hip (e) knee and (f) ankle}
	\label{fig:fig4}
\end{figure*}
\section{RESULTS}
We collected six joint angle profiles: elbow, shoulder, lumbar spine, hip, knee, and ankle. The joint angle profiles are shown in  Figure 4. We collected only the sagittal plane joint angle profiles. Same joint angles and GRF were collected by another group using a high precision multi-camera motion capture system and laboratory-grade force platform \cite{b4}.  There are some deviations for elbow joint angle profiles between Figure 4(a) and \cite{b4}. The trough of the elbow joint angle profile is around 90\% of the lifting cycle in Figure 4(a), whereas it is reported at 60\% of the lifting cycle in \cite{b4}. The reason for this deviation is the lifting weight. As the lifting weight  in \cite{b4} was heavy, the subject tried to keep the load and elbow close to the body for balancing at the middle stage of the lifting cycle. As a result, the trough of the elbow joint angle profile was 60\% of the lifting cycle. On the other hand, the lifting load for our subject was lighter than \cite{b4}. It was easy for the subject to maintain balance without keeping the elbow close to the body at the early stage. The pattern and phase change of the shoulder, lumbar spine, hip, knee and ankle  joint angle profiles in Figure 4 are similar to \cite{b4}. The total vertical GRF profile is shown in Figure 5. The maximum vertical GRF in Figure 5 is around 450N, whereas it was reported 1125N in \cite{b4}. That is normal, as the lifting weight for \cite{b4} was higher than our experiment. Except that, the pattern of the GRF in Figure 5 is similar to \cite{b4}.
\begin{figure}[htbp]
	\centering
	\includegraphics[scale=0.50]{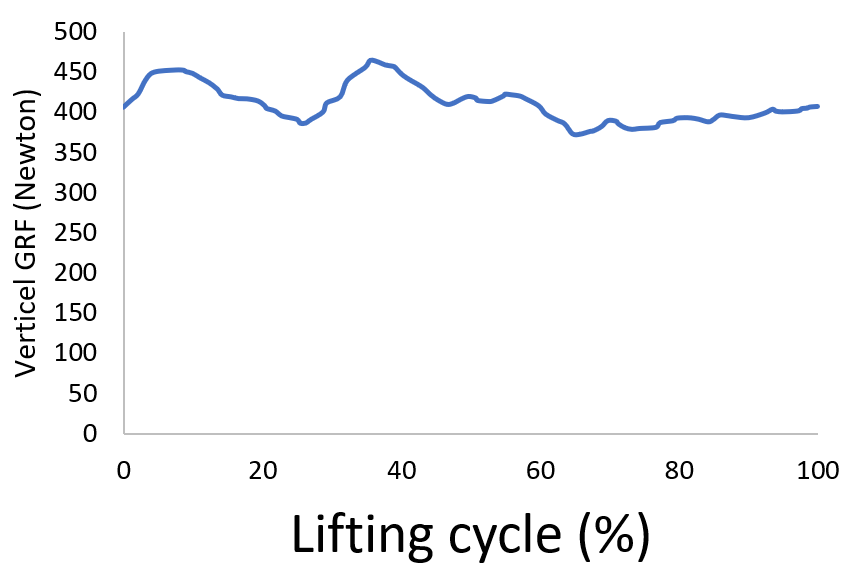}
	\caption{Total ground reaction forces}
	\label{fig:fig5}
\end{figure}\\
Using the static optimization tool in OpenSim, muscle forces data were calculated for two lower limb muscles: gluteus maximus and vastus intermedius (Figure 6). The maximum force generated from Vastus intermedius (Figure 6(b)) was 1702N. The location of the peak muscle force for vastus intermedius was at 40\% of the lifting cycle, whereas in \cite{b16} it was reported 10\% of the lifting cycle.  The lifting load reported in \cite{b16} was for maximum weight. The difference between peak points shows that different persons can adopt different strategies for lifting lightweight and heavyweight. The maximum muscle force for the glutei muscle was 1580N at about 80\% of the lifting time (Figure 6(a)).  The pattern of glutei muscle force is similar to the muscle force in \cite{b16}.

\section{Discussion and conclusion}
This study aims to develop a low cost and portable module to identify and assess MMH related injuries for developing countries.  To do this, we develop a comprehensive biomechanical analysis system to evaluate MMH related injuries. The joint angle profiles are collected using a smartphone camera and Kinovea. The vertical GRF profile is collected using Wii balance board and Brainblox software. Finally, the joint angles and GRF are used as the input in the OpenSim static optimization tool to find the muscle forces. The hardware used here are affordable and portable. The software used here are open-source and easy to use. The overall framework gives us access from collecting joint angle profiles to muscle force profiles during lifting weight.\\ 
The joint angle profiles collected using smartphone camera and Kinovea software are similar to literature \cite{b4} except elbow joints. The GRF profile is also agreed well with the literature \cite{b4}. The muscle force profiles from OpenSim static optimization tool are also similar to \cite{b16}.\\
\begin{figure}[h!]
	\centering
	\includegraphics[scale=0.7]{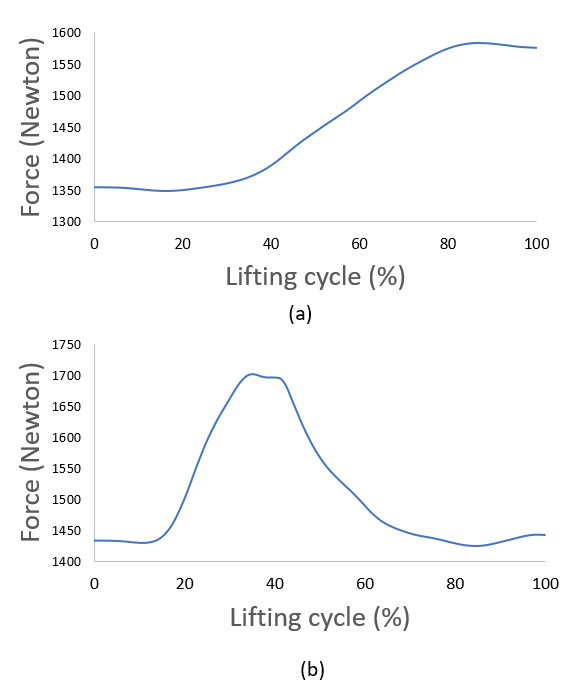}
	\caption{Lower limb muscle force for (a) gluteus maximus, (b) vastus intermedius}
	\label{fig:fig6}
\end{figure} 
Although this low cost biomechanical analysis framework shows some promising results, there are some limitations. First, this system can read only symmetric lifting motion. To study asymmetric lifting motion, two cameras are required. Moreover, lifting motions from both cameras need to be synchronized in Kinovea before extracting joint angle profiles. Secondly, there are some potential inaccuracies when synchronizing Kinovea and Wii balance board data. Our future plan is to develop a custom program to collect and process joint angle and GRF data simultaneously using same platform. This platform will reduce the synchronization error. Thirdly, Wii balance board can collect only vertical GRF data. It can not study horizontal GRF data. However, the horizontal GRF during lifting is insignificant.  Fourth, only one subject was used to validate the framework. Although the accuracy of the Wii balance board, Kinovea and OpenSim has been studied separately in several studies, further research should examine the accuracy of the overall system.\\
The construction and automobile repair industries in developing countries are mainly dependent on MMH related tasks. Multi-camera, force plate and electromyography system are not replaceable with this low cost platform for clinical use. But, the accuracy of this low cost portable system is adequate enough to evaluate MMH related injuries for different industrial workers in developing countries. In addition, this framework can be used for low funded research institutions, physiotherapists and sports injury analysis.

\vspace{12pt}
\color{red}


\begin{thebibliography}{00}
\bibitem{b1} J. N. Katz, ‘Lumbar disc disorders and low-back pain: socioeconomic factors and consequences’, J. Bone Joint Surg. Am., vol. 88 Suppl 2, pp. 21–24, Apr. 2006, doi: 10.2106/JBJS.E.01273.
\bibitem{b2} J. Song, X. Qu, and C.-H. Chen, ‘Simulation of lifting motions using a novel multi-objective optimization approach’, Int. J. Ind. Ergon., vol. 53, pp. 37–47, 2016.
\bibitem{b3} T. R. Waters and A. Garg, ‘Two-dimensional biomechanical model for estimating strength of youth and adolescents for manual material handling tasks’, Appl. Ergon., vol. 41, no. 1, pp. 1–7, Jan. 2010, doi: 10.1016/j.apergo.2009.02.006.
\bibitem{b4} Y. Xiang, J. Cruz, R. Zaman, and J. Yang, ‘Multi-objective optimization for two-dimensional maximum weight lifting prediction considering dynamic strength’, Eng. Optim., p. (in press), 2020.
\bibitem{b5} Y. Xiang, R. Zaman, R. Rakshit, and J. Yang, ‘Subject-specific strength percentile determination for two-dimensional symmetric lifting considering dynamic joint strength’, Multibody Syst. Dyn., vol. 46, no. 1, pp. 63–76, May 2019, doi: 10.1007/s11044-018-09661-1.
\bibitem{b6} R. Zaman, Y. Xiang, J. Cruz, and J. Yang, ‘Two-dimensional versus three-dimensional symmetric lifting motion prediction models: a case study’, ASME J. Comput. Inf. Sci. Eng., p. (accepted), 2020.
\bibitem{b7} Y. Xiang, J. S. Arora, and K. Abdel-Malek, ‘3d human lifting motion prediction with different performance measures’, Int. J. Humanoid Robot., vol. 09, no. 02, p. 1250012, Jun. 2012, doi: 10.1142/S0219843612500120.
\bibitem{b8} Y. Xiang, J. S. Arora, S. Rahmatalla, T. Marler, R. Bhatt, and K. Abdel-Malek, ‘Human lifting simulation using a multi-objective optimization approach’, Multibody Syst. Dyn., vol. 23, no. 4, pp. 431–451, Apr. 2010, doi: 10.1007/s11044-009-9186-y.
\bibitem{b9} R. Zaman, Y. Xiang, J. Cruz, and J. Yang, ‘Three-dimensional asymmetric maximum weight lifting prediction considering dynamic joint strength’, Proc. Inst. Mech. Eng. [H], p. (accepted), 2020.
\bibitem{b10} R. Zaman, Y. Xiang, J. Cruz, and J. Yang, ‘Three-dimensional symmetric maximum weight lifting prediction’, St. Louis, MO, Aug. 2020, p. (in press).
\bibitem{b11} S. L. Delp et al., ‘OpenSim: open-source software to create and analyze dynamic simulations of movement’, IEEE Trans. Biomed. Eng., vol. 54, no. 11, pp. 1940–1950, 2007.
\bibitem{b12} E. Beaucage-Gauvreau et al., ‘Validation of an OpenSim full-body model with detailed lumbar spine for estimating lower lumbar spine loads during symmetric and asymmetric lifting tasks’, Comput. Methods Biomech. Biomed. Engin., vol. 22, no. 5, pp. 451–464, 2019.
\bibitem{b13} Y. Blache, F. Dal Maso, L. Desmoulins, A. Plamondon, and M. Begon, ‘Superficial shoulder muscle co-activations during lifting tasks: Influence of lifting height, weight and phase’, J. Electromyogr. Kinesiol., vol. 25, no. 2, pp. 355–362, 2015.
\bibitem{b14} Y. Blache, L. Desmoulins, P. Allard, A. Plamondon, and M. Begon, ‘Effects of height and load weight on shoulder muscle work during overhead lifting task’, Ergonomics, vol. 58, no. 5, pp. 748–761, 2015.
\bibitem{b15} G. M. R. uz zaman Rana, ‘Maximum weight lifting prediction considering dynamic joint strength’, Thesis, 2018.
\bibitem{b16} R. Zaman, Y. Xiang, R. Rakshit, and J. Yang, ‘Muscle Force Prediction in OpenSim Using Skeleton Motion Optimization Results As Input Data’, in Volume 1: 39th Computers and Information in Engineering Conference, Anaheim, California, USA, Aug. 2019, p. V001T02A046, doi: 10.1115/DETC2019-97520.
\bibitem{b17} R. M. Abd El-Raheem, R. M. Kamel, and M. F. Ali, ‘Reliability of using Kinovea program in measuring dominant wrist joint range of motion’, Trends Appl. Sci. Res., vol. 10, no. 4, p. 224, 2015.
\bibitem{b18} Fernández-González, A. Koutsou, A. Cuesta-Gómez, M. Carratalá-Tejada, J. C. Miangolarra-Page, and F. Molina-Rueda, ‘Reliability of Kinovea® Software and Agreement with a Three-Dimensional Motion System for Gait Analysis in Healthy Subjects’, Sensors, vol. 20, no. 11, p. 3154, 2020.
\bibitem{b19} C. Guzmán-Valdivia, A. Blanco-Ortega, M. Oliver-Salazar, and J. Carrera-Escobedo, ‘Therapeutic motion analysis of lower limbs using Kinovea’, Int J Soft Comput Eng, vol. 3, no. 2, pp. 2231–307, 2013.
\bibitem{b20} N. A. H. Hisham, A. F. A. Nazri, J. Madete, L. Herawati, and J. Mahmud, ‘Measuring ankle angle and analysis of walking gait using kinovea’, 2017, pp. 247–250.
\bibitem{b21} Parida, R., and Ray, P. K., 2015, Biomechanical modelling of manual material handling tasks: A comprehensive review. Procedia Manufacturing, 3, 4598-4605.
\bibitem{b22} A. Puig-Diví, J. M. Padullés-Riu, A. Busquets-Faciaben, X. Padullés-Chando, C. Escalona-Marfil, and D. Marcos-Ruiz, ‘Validity and reliability of the kinovea program in obtaining angular and distance dimensions’, 2017.
\bibitem{b23} R. A. Clark, A. L. Bryant, Y. Pua, P. McCrory, K. Bennell, and M. Hunt, ‘Validity and reliability of the Nintendo Wii Balance Board for assessment of standing balance’, Gait Posture, vol. 31, no. 3, pp. 307–310, 2010. 
\bibitem{b24} J. M. Leach, M. Mancini, R. J. Peterka, T. L. Hayes, and F. B. Horak, ‘Validating and calibrating the Nintendo Wii balance board to derive reliable center of pressure measures’, Sensors, vol. 14, no. 10, pp. 18244–18267, 2014.
\bibitem{b25} Littrell, M. E., Chang, Y.-H., and Selgrade, B. P., 2018, Development and assessment of a low-cost clinical gait analysis system. Journal of applied biomechanics, 34, 503-508.
\bibitem{b26} Yang, C. X., 2018, Low-cost experimental system for center of mass and center of pressure measurement (june 2018). IEEE Access, 6, 45021-45033.
\bibitem{b27} Chakravarty, K., Chatterjee, D., Das, R. K., Tripathy, S. R., and Sinha, A., 2017, Analysis of muscle activation in lower extremity for static balance. 2017 39th Annual International Conference of the IEEE Engineering in Medicine and Biology Society (EMBC), 4118-4122.
\bibitem{b28} Cooper, J., Siegfried, K., and Ahmed, A., 2014, Brainblox: Brain and biomechanics lab in a box software. Version.
\bibitem{b29} Rajagopal, A., Dembia, C. L., DeMers, M. S., Delp, D. D., Hicks, J. L., and Delp, S. L., 2016, Full-body musculoskeletal model for muscle-driven simulation of human gait. IEEE Transactions on Biomedical Engineering, 63, 2068-2079.
\end{thebibliography}
\end{document}